# Electrocatalytic CO2 Reduction: Role of the Cross-Talk at Organic-Inorganic Interfaces

Michele Melchionna *[a], Paolo Fornasiero [ab], Maurizio Prato [acd], Marcella Bonchio *[e]

[a] Department of Chemical and Pharmaceutical Sciences, Center for Energy, Environment and Transport Giacomo Ciamician, University of Trieste and Consortium INSTM, Via L. Giorgieri 1, 34127 Trieste, Italy

[b] ICCOM-CNR, University of Trieste, Via L. Giorgieri 1, 34127 Trieste, Italy

[c] Center for Cooperative Research in Biomaterials (CIC biomaGUNE), Basque Research and Technology Alliance (BRTA), Paseo de Miramón 194, Donostia San Sebastián, Spain

[d] Ikerbasque, Basque Foundation for Science, Bilbao, Spain

[e] Department of Chemical Science, ITM-CNR, University of Padova and Consortium INSTM, Via F. Marzolo 1, 35131 Padova, Italy

E-mail: melchionnam@units.it, marcella.bonchio@unipd.it

**Abstract**

A Electrocatalytic $CO_2$RR is an interfacial process, involving a minimum of three phases at the contact point of gaseous $CO_2$ with the electrodic surface and the liquid electrolyte. As a consequence, surface chemistry at composite interfaces plays a central role for $CO_2$RR selectivity and catalysis. Each interface defines a functional boundary, where active sites are exposed to a unique environment with respect to distal sites in the bulk or organic and inorganic materials. While the individual role of each component-type is hardly predictable "a-solo", the interface ensemble works via a strategic interplay of individual effects, including: (i) enhanced electrical conductivity, (ii) high surface area and exposure of the interfacial catalytic sites, (iii) favorable transport and feeding of reactants, (iv) complementary interactions for the "on/off" stabilization of cascade intermediates, (v) a secondary sphere assistance to lower the activation energy of bottleneck steps, (vi) a reinforced robustness and long-term operation stability by mutual protection and/or healing mechanisms. Selected $CO_2$RR case studies are compared and contrasted to highlight how the organic domains of carbon nanostructures merge with metal and metal-oxide active sites to separate tasks but also to turn them into a cooperative asset of mutual interactions, thus going beyond the classic "Divide et Impera" rule.

**Broader context**

"Modern civilization is the daughter of coal": this is Ciamician's opening sentence in his 1912 visionary Science paper. Indeed carbon is the primary component of our Life, as we know it, of all organic matter in our body, food, gasoline, drugs and in a million solid/liquid/gas chemicals that we use routinely. All these chemicals are constantly produced by fossil carbon sources, while it takes hundred million years to restore fossil feedstock by the so called “slow carbon cycle” over rocks, soil, ocean, and atmosphere. The way we consume fossil carbon is too intensively altering the balance of the carbon cycle on Earth and putting our future at risk. The alternative is to intercept the natural carbon cycle at the $CO_2$ stage, implementing a synthetic “fast carbon cycle” using $CO_2$ as abundant, ubiquitous, C1-reagent for the next generation chemical industry. Electrocatalytic $CO_2$ reduction offers an appealing perspective especially considering the combined application of solar photovoltaics and renewable sources for electrical power generation. One major advantage is that $CO_2$ reduction products and intermediates can be processed and exploited within already existing infrastructures and chemical plants. The goal ahead is to translate the actual frontier research on $CO_2$RR into the market, which means a huge effort dedicated to increase efficiency, selectivity and scaled-up catalytic methods in order to be competitive with fossil-fuelled production, reduce our carbon foot-print and accelerate the most desirable transition to a net zero-carbon economy.

Inspired by Nature, conversion of $CO_2$ into added-value chemicals needs a complex synthetic machinery, regulated by a most effective confinement of reagents, task-separation, orchestration of rates and functions by making extensive use of specialized interfaces and hybrid organic-inorganic domains for biological $CO_2$ processing. The expectation for the next generation electrocatalyst is to rival the natural asset, through a creative design of functional interfaces and new contamination across scientific disciplines. The vision is to merge materials science and tailored electrocatalytic interfaces with biological routines. Taking the best of the two worlds, by coupling artificial $CO_2$RR with biological $CO_2$ fixation. “What is next is great and breathtaking”, as the new president Joe Biden said about the future of Science: we know it is our responsibility.

## 1. Introduction.

Under the Paris Agreement, the United Nations took responsibility for the control of global warming thus counteracting the risks of climate change. This priority action calls into play any possible strategy for $CO_2$ abatement, to "achieve a balance between anthropogenic emissions by sources and removals by sinks of greenhouse gases". (Paris Agreement 5$^{th}$ October 2016, COP21)

The current strategic plan for $CO_2$ mitigation contemplates several approaches, among which two are expected to be highly promising (Figure 1). The "$CO_2$ capture and storage" approach (CCS) is based on sequestration of gaseous $CO_2$ by absorbing materials, that can mineralize $CO_2$ to carbonates.[1] However, CCS presents the main issue of $CO_2$ long-term storage safety and stability.[2] On the other hand, the "$CO_2$ chemical fixation" approach (CCF), recycling of $CO_2$ into valuable carbon-containing products, offers bright horizons considering: (i) the most convenient storage of liquid $CO_2$-derived products at ambient conditions; (ii) their potential as renewable combustion fuels, powering an overall carbon-neutral energy cycle;[3, 4] (iii) the added value of a circular atom economy scheme, where key commodity chemicals can be produced from $CO_2$ as the C1-buliding block via its selective reduction into $HCO_2H$, $CH_3OH$, $CH_4$, and/or C-C coupling products. The $CO_2$ reduction reaction (generally referred to as "$CO_2$RR") can be performed with different methods including photo-, electro-, thermal and enzymatic catalysis.[5] In particular, any fundamental progress on the electrochemical $CO_2$ processing is central to the development of new electro-enzymatic and photo-electrocatalytic schemes, which are gaining increasing attention both from a mechanistic and a synthetic perspective.

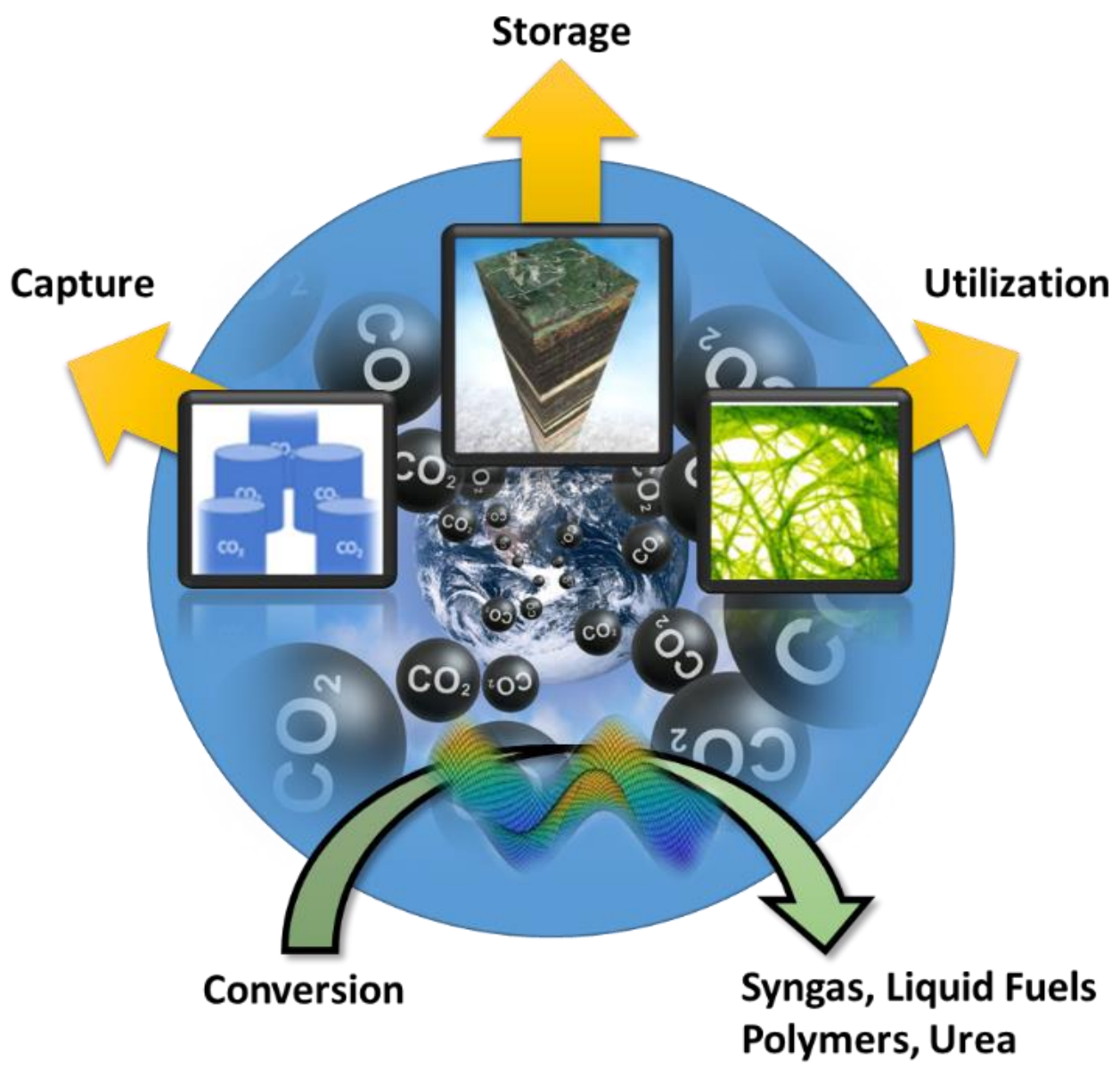

**Figure 1.** $CO_2$ mitigation approaches highlighting the conversion of $CO_2$ into valuable carbon-containing products, via chemical fixation through reaction pathways carved on tailored potential energy surfaces.

Electrochemical $CO_2$RR, especially if carried out in aqueous phase, is amenable to clean energy schemes and sustainable "green-chemistry" methods, when renewable sources (i.e. solar photovoltaics) are used to generate the required electrical potential, and considering mild catalytic electrohydrogenation conditions that can avoid a direct $H_2$ supply, at high pressure and high temperature conditions. Compared to direct photo-reduction protocols operating upon a photo-induced charge separation and dictated by the photophysical properties of the photoactive materials,[6] "dark" electrochemical $CO_2$RR offers the great advantage of tuning the applied potentials ($E_{ap}$), according to the kinetic and thermodynamic requirements of the selected reaction (overpotential), thus avoiding competitive pathways and favouring selectivity. Moreover, the gap between the fundamental progress on electro-catalyst development and the technical hurdles for device implementation is expected to be bridged in shorter time frame as compared to other less-mature technologies.[7]

The grand challenge of electrochemical $CO_2$RR lies in the design of next-generation electrocatalysts that can sustain a high current efficiency at low overpotential ($\eta = E_{ap} - E^0$, namely the potential to be applied that exceeds the equilibrium potential, $E^0$), while featuring a favorable selectivity towards target products, long term stability and sustainable cost associated to the any synthetic protocol, scale-up and recycling. The number of proposed functional molecules and materials is incessantly increasing. Among these latter, the synthesis of multi-phase, hybrid nano-materials is gaining a big momentum with the precise ambition to control the multi-component structure, morphology and hierarchy of the final composite, while addressing the specific functions of the redox-active core, of secondary-sphere interactions and of relevant interfacial phenomena. With this aim, the engineering of functional organic-inorganic nano-hybrids for $CO_2$RR has the potential to set a new paradigm in the field of electro-catalysis for multi-redox transformations and small molecule activation.[8-10] The main problem with $CO_2$ reduction lies in its high chemical inertness so that $CO_2$RR generally proceeds through a complex proton coupled multi-electron mechanism, dictated by both thermodynamic restrictions and kinetic hurdles. The purpose in the synthetic design of hybrid nanomaterials is to bring up synergistic effects that can orchestrate $CO_2$RR by favoring a cooperative interplay of absorption and confinement effects, multi-site across-boundary reactivity, interfacial dynamics affecting the kinetic of each elementary step and giving access to low-energy mechanistic pathways. Inspiration is drawn from the complexity of $CO_2$RR in biological systems occurring at equilibrium potential and performed by specialized enzymatic machineries.[4,11] However, artificial analogs of $CO_2$ reduction enzymes are still far from the biological performance that can be compared under electrocatalytic conditions in terms of overpotential, current density (*j*, the measured current divided by the geometric surface area of the working electrode), turnover frequency (TOF), faradaic efficiency (FE, the fraction of consumed charge actually used in the conversion to a given product), selectivity and long-term stability. Significant advancements have been made, considering bio-inspired functional guidelines to shape the electrocatalytic machinery , while avoiding a mere replica of the energy-intensive biological structure.[12] This implies that man-made building blocks and their functional assembly will be optimized to counteract both the intrinsic fragility of natural proteins and catalytic co-factors and their high-energy processing within the enzyme active sites. Therefore, the roadmap to shape artificial multi-redox routines for efficient $CO_2$ activation will require the evolution of organic-inorganic hybrid conjugates, displaying multi-phase catalytic domains that are amenable to modular architectures with the aim to control: (i) the composition of the diverse domains at the atomic level (including structural defects, hetero-dopants, terminal groups, redox manifold etc..) (ii) the surface/interface engineering of sub-domain boundaries; (iii) the overall morphology and hierarchical phase arrangement; (iv) the reactive sites distribution, their phase-segregation and/or inner-sphere contacts; (v) any competent second sphere interactions emerging from the active sites surrounding; (vi) inter-phase transport dynamics and intermediate stabilization.[10] Here, we discuss the critical points connected with latest progresses on composite electrocatalysts for enhanced $CO_2$ reduction particularly focusing on hybrid nanomaterials by dissecting the role of (sub)-structures and identifying the new functional capacity of the ensemble. A perspective on emerging research directions is highlighted in the conclusion section of the manuscript.

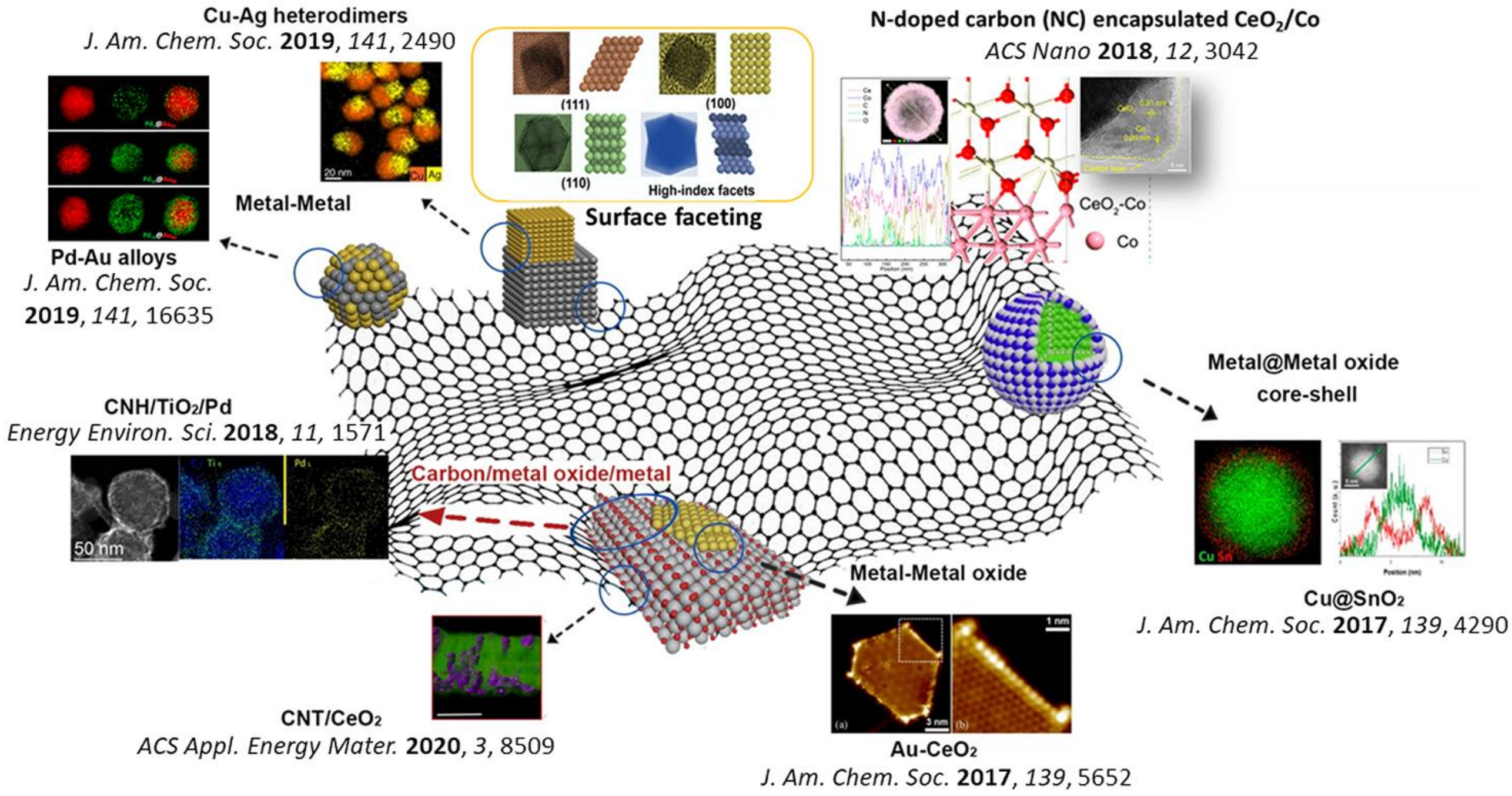

**Figure 2.** Types of typical material interfaces used in electrocatalysis. Adapted with permissions from ref. 28 (Cu-Ag heterodimers), 29 (Pd-Au alloys), 30 (N-doped carbon/$CeO_2$/Co, 50 (Au-$CeO_2$), 55 (Cu@$SnO_2$), 71 (CNH/$TiO_2$/Pd), 72 (CNT/$CeO_2$).

## $CO_2$RR Electrocatalysis: general aspects.

Enhancement of electrocatalytic performance has been primarily pursued by focusing on two aspects: 1) optimization of the intrinsic activity of active sites and 2) increasing the number of available active sites. This dual approach entails a rational effort based on combined theory and modelling descriptors with related experimental evidence. The aim is to convey a full set of fundamental principles regulating electrocatalytic transformations with broad application and by a multi-level analysis of the catalytic performance.[13] However, $CO_2$ reduction ($CO_2$RR) poses some unique challenges, compared to other small molecule activation and energy-related reactions including: oxygen reduction (ORR), hydrogen evolution (HER) water oxidation (WOR), or nitrogen reduction(NRR).

The complexity of $CO_2$RR is by far related to the great number of possible products that can be generated under electrocatalytic conditions, implying diverse mechanistic steps and/or consecutive transformations with specific thermodynamic and kinetic requirements, including the competitive HER occurring in protic media.[14, 15]

The selectivity issue calls for attention, as any benefit arising from increasing of the electrocatalyst performance might be overriden by a selectivity loss, due to a poor control over the diverse reaction coordinates leading to multiple products formation.[16, 17] The formation of the radical anion $CO_2^-$ by the first electron reduction occurs at very negative potentials, due to the large structural reorganization of the bent radical anion ($E_0$ = -1.90 V, vs SHE in an aqueous solution, pH 7).[17] This step stands as the rate-determining step preceeding a multi-step reduction sequence in $CO_2$RR. In this regard, the thermodynamic potential of proton reduction (HER process) at pH 7 ($E_0$ = -0.42 V, vs SHE) occurs at less negative potentials than the $CO_2^{\cdot -}$ radical anion formation. Overall, compared to $CO_2$RR, HER turns out to be favored when operating in proton-rich electrolytic solutions, such as aqueous media. This generally leads to low faradaic efficiency (FE) for the desired $CO_2$ reduction product, as most of the transferred electrons are used to generate $H_2$. Several strategies can be adopted to overcome this problem:

(i) the use of aprotic solvents and electrolytes to suppress HER;

(ii) a tailored engineering of the catalyst package in terms of its atomic-scale structure, surface and interfacial properties that favor $CO_2$ absorption and transport while increasing the overpotential gap for water/proton reduction;[18]

(iii) the fabrication of porous, mesostructured electrodes, to impact wettability and proton diffusion so to inhibit HER while favoring $CO_2$ enrichment at catalytic sites.[19]

Indeed, the interplay of all these effectors are instrumental to control the $CO_2$RR selectivity outcome. Concerning non-aqueous electrolytes, ionic liquids (ILs) represent a greener alternative to organic solvents, for $CO_2$ solubilization and for stabilization of charged reduction intermediates.[20-24] However, due to cost issues, the scale-up of electrochemical devices using IL-based electrolytes is not straightforward. In this respect, the recent application of deep eutectic solvents (DESs) for $CO_2$RR offers a promising perspective. DESs are usually binary/ternary mixtures of hydrogen bond donor/acceptor molecules, whose melting points are substantially lower than those of the separated components, thus exhibiting low vapor pressure, high conductivity, a wide electrochemical potential window, and high $CO_2$ solubility, as conventional ILs. The advantage of DESs, is their ability to significantly decrease the onset potential for the $CO_2$RR, by favoring the proton-coupled electron transfer (PCET) mechanism, which is instrumental to enhance selectivity. As in the case of choline-based DESs, these systems are generally nontoxic and less expensive than ILs, while serving as additive phases and/or organo-catalysts for the elecrocatalytic $CO_2$RR.[25]

The tailored choice of multi-phase catalytic domains is indeed one emerging strategy to target the HER $\rightarrow$ $CO_2$RR selectivity switch,[26] regulated by the applied potential. Ideally, the design of a hybrid, organic-inorganic, catalytic interface allows to by-pass the first electron injection step (formation of the $CO_2$ radical anion) favoring alternative mechanisms via the stabilization of diverse $CO_2$-based intermediates. The result is a definite shift of the $CO_2$RR onset potential at earlier potentials. This was shown by Kanan and Min, who used Pd nanoparticles supported on carbon for the direct electro-hydrogenation of $CO_2$ to formic acid, at near equilibrium potentials so ruling out the high-energy formation of the radical anion.[27]

## Shaping Multi-Phase Interfaces for Bio-inspired $CO_2$RR Electrocatalysis

Electrocatalytic $CO_2$RR is an interfacial process,[28-30] involving a minimum of a triple phase boundary at the contact point of gaseous $CO_2$ with the electrodic surface and the liquid electrolyte. Indeed, interfacial $CO_2$RR takes place through sequential $CO_2$ adsorption, surface diffusion and activation at catalytic sites, and the ultimate step of product desorption. Because $CO_2$ transport and accumulation is dictated by favourable equilibria at the catalytic sites, the reaction performance depends on the density of the active centers and by proximal cooperative effects emerging from the catalyst local environment and morphology. Therefore, the design of $CO_2$RR electrocatalysts is today flourishing in the field of multi-phase materials, where optimized interfaces hold the key for enhanced catalysis, regulating the stereo-electronic requirements of the active sites, a high interfacial-to-bulk ratio of their distribution, and together with porous architectures favoring the access of both $CO_2$ and of the liquid electrolyte.

Types of interfaces under the lens include not just binary or ternary metal junctions, but heterojunctions with metal/metal oxide and hybrid organic-inorganic interfaces, where a carbon-based framework is intimately connected to metal/metal oxide domains with the overall effect to provide combined kinetic and thermodynamic advantages.[31]

From a mere stability perspective, it is of general knowledge that metal oxides and carbon-based scaffolds can improve dispersion and stability of noble metal nanoparticles, and even that of single atom catalytic sites.[32] Indeed, oxide-based materials are commonly used as robust heterogeneous supports for industrial catalytic applications.[33]

With respect to electrocatalysis, the added-value of the composite metal oxide phase and of the organic hetero-junctions is their role in providing multifunctionality thus responding to fundamental requirements of the $CO_2$RR mechanistic envelope. This aspect is of particular relevance for the $CO_2$RR selectivity issue, that can take advantage from specific functions emerging from the diverse catalytic domains while being integrated in one single electrocatalytic platform. Indeed, selective $CO_2$RR stems from a stringent control on diverse functional steps occurring in a parallel or cascade mode, that dictate the dominant reduction pathway and the product distribution. Table 1 collects the multifunctionality requirements for $CO_2$RR considering the final product distribution and selected electrocatalytic active sites associated to these functions.

**Table 1**. Multifunctionality of the $CO_2$RR electrocatalyst classified according to the dominant product distribution and mechanistic pathway.[34]

| **$CO_2$RR main products** | **Key Functional steps** | **Selected active sites** |
|---|---|---|
| CO | (i) binding to form a carboxylic acid intermediate (*COOH)<br>(ii) low binding energy of the *CO intermediate | Ag, Au and Zn metal sites, single-atom Fe/Ni sites[18, 35-37] |
| HCOOH | (i) one electron reduction to $CO2^{-\bullet}$ radical<br>(ii) protonation to form the *OCHO intermediate | Pb, Hg, Sn, Bi metal sites[38-41] |
| HCOOH electro-hydrogenation | (i) formation of reactive hydrides (M-H) followed by<br>(ii) CO2 insertion to form *COOH | Pd,[27] |
| $CH_3OH$ or favorable C-C couplings | (i) moderate/strong *CO binding energy allowing a cascade reduction events<br>(ii) stabilization of *CHO and *OCCO intermediates and possibility to form | Cu, $Cu_2O$ [42-44] |
| Suppression of $H_2$ evolution (HER) | (i) weak binding energy with *H<br>(ii) favorable $CO_2$ adsoption and diffusion | Fe/Ni, Au defects[45] |

Therefore, the $CO_2$RR selectivity is governed by the relative energies of $CO_2$ binding modes, and of key reduction intermediates at the catalytic surface, which are mainly regulated by the metal-site ability of electron back donation from its d orbitals (Table 1). Hence, the engineering of the active site stereo-electronic features is expected to control selectivity by giving access to reaction pathways at lower energy cost. The new paradigm emerging from these observations is the central importance of surface chemistry to address selective catalysis by the rationale assembly of composite interfaces. As a general concept, each interface defines a functional boundary, where active sites are exposed to a unique environment that differs from that of bulk distal sites. The chemical and electronic properties of interfacial sites can thus be exploited to leverage effective catalysis. *As a consequence, reactivity and selectivity are tuned at diverse functional interfaces, that can be instrumental to separate tasks but also to merge into a cooperative asset of mutual interactions, thus going beyond the classic "Divide et Impera" rule*.

This strategy is extensively adopted by Nature, as surfaces and interfaces are often the preferential frameworks to accomplish vital, but difficult, biological processes. The same approach can therefore be translated within artificial architectures that can be designed for especially demanding catalytic applications.

For $CO_2$ electro-catalysis, the concept of *"collaborative catalytic interfaces"* was proved at least a decade ago, when Hori et al. while working on pure Sn, a known catalyst for electrogeneration of HCOOH from $CO_2$,[46] noted that the formation of a $SnO_x$ native layer on the Sn electrode resulted in an 8-fold increase in current density and a 4-fold increase in HCOOH production, measured as faradaic efficiency (FE). In contrast, the removal of the $SnO_x$ layer reverted the catalysis to HER, revealing the key role of the metal oxide component for CO2RR selectivity. It was assumed that $SnO_x$ could either stabilize the incipient negative charge on $CO_2$ or could act as an electron transfer mediator.[47]

*While the individual role of each component is hard to be detected "a-solo", the overall engineering of multifunctional electrocatalytic interfaces is meant to provide a combined cross-talk of individual effects*, including: (i) enhanced electrical conductivity, (ii) high surface area and exposure of the interfacial catalytic sites, (iii) favorable transport and feeding of reactants, (iv) complementary interactions for the "on/off" stabilization of cascade intermediates (Table 1), (v) a secondary sphere assistance to lower the activation energy of bottleneck steps, (vi) a reinforced robustness and long-term operation stability by mutual protection and/or healing mechanisms. The interplay of these interfacial properties offers a wide space of exploration under electrocatalytic conditions. A fundamental tool is certainly provided by the continuous advancement of specialized characterization techniques (operando spectroscopies aided by computational studies) to pinpoint the intimate features of the electrocatalyst structure and of the multi-phase arrangement evolving under catalytic regime.

Certainly $CO_2RR$ at copper sites is highly promising, on account of Cu wide availability on Earth and with regard to its privileged selectivity favoring $C_{\geq 2}$ hydrocarbons. Higher hydrocarbons, with higher energy density than $C_1$ products, are versatile feed-stocks and generally obtained by petroleum refining or by Fischer–Tropsch synthesis with $H_2$ and high temperature conditions. A direct electrocatalytic production of higher hydrocarbons from $CO_2$ requires a complex multi-electron/multi-proton transfer mechanism and the formidable challenge of forming new C-C bonds. Therefore, the choice of a metal-oxide composite to boost $CO_2RR$ at copper sites has been considered a valuable strategy.

To this aim, $CeO_2$ is being considered for $CO_2$RR due to its rich redox chemistry associated with a dynamic evolution of oxygen vacancies under electrocatalytic conditions. These properties are expected to induce additional binding states of key intermediates and direct the $CO_2$RR selectivity.[48] As a representative example, Cu/$CeO_{2-x}$ nanocrystalline heterodimers have been reported to effect the conversion of $CO_2$ to $CH_4$ with FE of 54 %.[49] In this case, the $CeO_{2-x}$ vacancies provide additional adsorption sites, that can stabilize $CO_2$RR intermediates by bidentate binding modes at adjacent Ce and Cu atoms. High resolution TEM was used to characterize the interfacial regions of the nanocomposite, that is consistent with an epitaxial connection between the ceria and the copper domains (Figure 3).

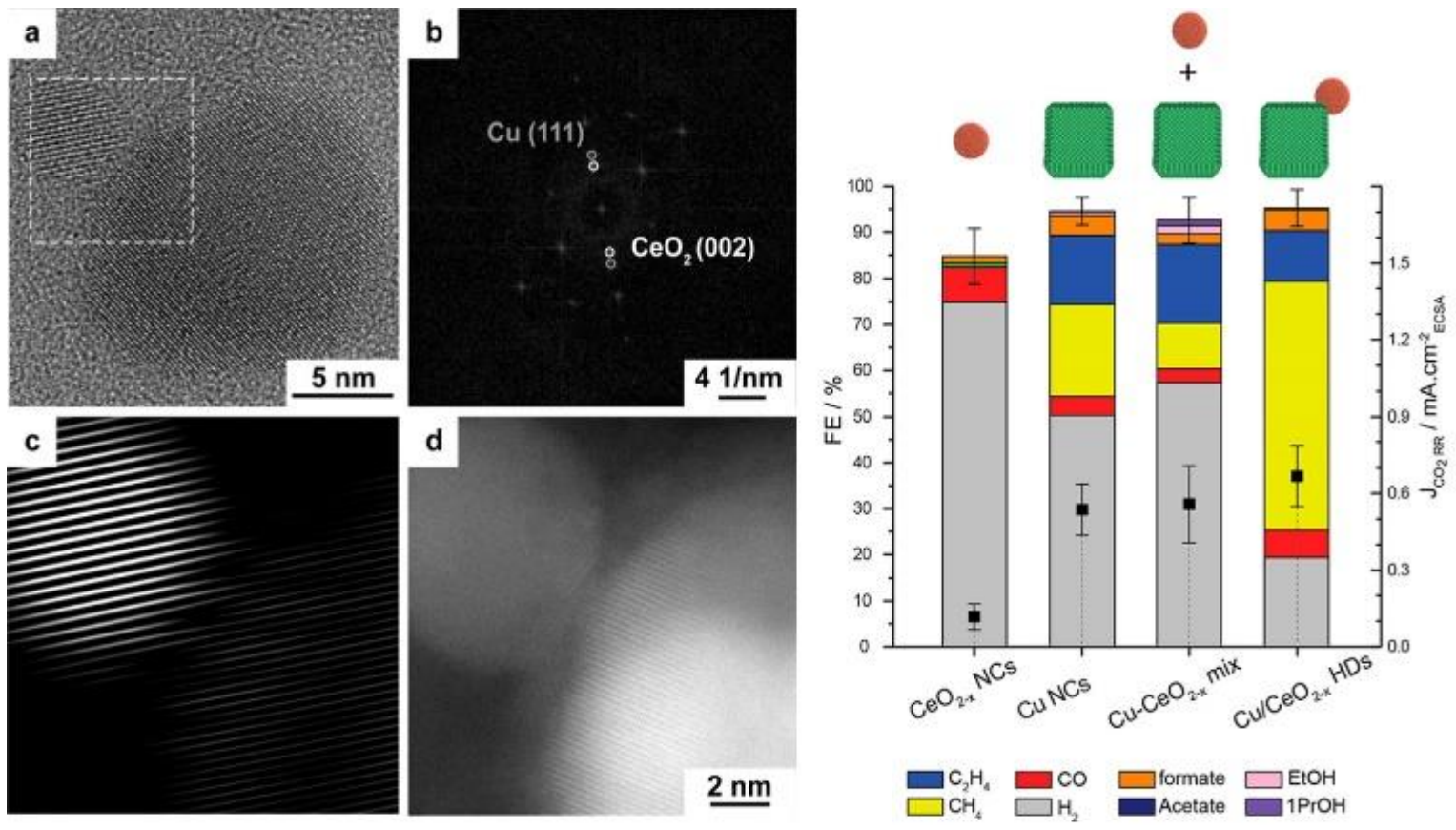

**Figure 3**. (a) HR-TEM image of one representative Cu/$CeO_{2-x}$ HD, (b) FFT diffractogram of the interfacial region, (c) inverse FFT (pseudo-dark field image), and (d) HR-STEM image of the interfacial region parts. Right panel: Faradaic efficiencies and CO2RR partial current-densities for 15 μg of Cu/$CeO_{2-x}$ HDs, Cu–$CeO_{2-x}$ mix, Cu NCs, and $CeO_{2-x}$ NCs loaded on a glassy carbon surface of 1 $cm^2$, measured at −1.2 $V_{RHE}$. Adapted with permissions from ref. 49. Copyright (2019) American Chemical Society.

Also in the case of Au and Ag electrocatalysts, the construction of the metal–$CeO_x$ interface leads to a significant enhancement of $CO_2$RR, that was not observed with the metal catalysts or with the metal-oxide phase alone. The CO Faradaic efficiency turns out to be > 89% over Au–

$CeO_x/C$ at −0.89 V vs reversible hydrogen electrode (RHE), which is higher than the sum of the individual performance of the single Au sites (59.0%) and of the $CeO_x$ phase (9.8%). The Au-$CeO_x$ interface is therefore essential for $CO_2$ absorption and diffusion at the electrocatalytic sites, as probed by synchrotron-radiation photoemission spectroscopy (SRPES), while $CO_2$ does not adsorb on Au surface even upon extended $CO_2$ exposure.[50] DFT calculations confirmed that the $CO_2RR$ active sites were located at the metal-metal oxide interface. Moreover, the electrocatalytic activity was found to depend on the percentage of reduced $Ce^{3+}$ sites, which is facilitated by redistribution of oxygen vacancies from bulk to surface.

Oxygen vacancies have been identified as one crucial feature in various other metal oxide-based electrocatalysts, by virtue of superficial charge modulation, which favorably alters $CO_2$ adsorption and activation.[51, 52]

Molecular metal oxides such as polyoxometalates (POM) represent a structurally defined component to be connected to metal phases. A remarkable observation was that highly challenging $CO_2RR$ product such as acetate could be formed with excellent FE (ca 49%) and a very high current density (∼110 mA $cm^{-2}$) by combining copper nanocubes with a molybdenum-based POM as the catalyst.[26] The outstanding performance originates from the interfacial Cu-O-Mo (confirmed by XAFS), whereby the Mo modifies electronically the Cu local structure to tune the product selectivity. According to DFT calculations, the key intermediate is $*CH_3$, which can favorably couple with $CO_2$ forming acetate with a lower energy profile. Small amounts of other products such as methane, ethylene and ethane, were observed depending on the fractional Cu surface not being covered with the Mo-contaning POM, thus highlighting the instrumental role of the Cu-O-Mo interface.[26]

**Hierarchical Metal@Metal-oxide Interfaces: the core-shell motif**

The assembly of hierarchical systems is attracting a considerable attention because of the multi-level arrangement of catalytic interfaces.[53, 54] In the realm of hierarchical structures, the core-shell motif emerges as an appealing choice for tailoring the catalyst properties. In particular, the interfacial confinement of metal nanoparticles (NPs) within a porous metal-oxide environment is expected to be crucial for selective $CO_2$RR. The metal oxide phase is instrumental considering a combination of favorable effects to enhance the $CO_2$RR, namely:

(i) promoting $CO_2$ adsorption at the porous nano-oxide architecture, can increase the concentration of $CO_2$ at the active sites, so to accelerate its conversion;

(ii) facilitated mass transport and gaseous product desorption at tailored metal-oxide surfaces can be a winning strategy to tune kinetics, tandem reactions and selectivity outcome;

(iii) the control of the local pH by buffering the acidity/basicity conditions after the electrocatalytic event can suppress the competitive HER;

(iv) control of the redox sites dynamics and oxygen vacancies at the interface will impact the $CO_2$ activation modes and the stabilization of reduction intermediates.

In a recent example, Cu nanoparticles enveloped within $SnO_2$ shells exhibited variable selectivity depending on the thickness of the oxide layer. The synergistic effect and cooperative phase interactions are demonstrated by the $CO_2$RR selectivity outcome, that turns out to depend on the core-shell relative dimensions: shell-free Cu NPs yield just small amounts of $C_2H_4$ and $C_2H_6$ while core-shell Cu@$SnO_2$ hybrids with thicker oxide layer (1.8 nm) lead to the prevalent formation of HCOOH (85% FE at −0.9 V). Noteworthy, the selective production of CO (93% FE) is achieved upon reducing the thickness of the tin-oxide shell (Figure 4).[55]

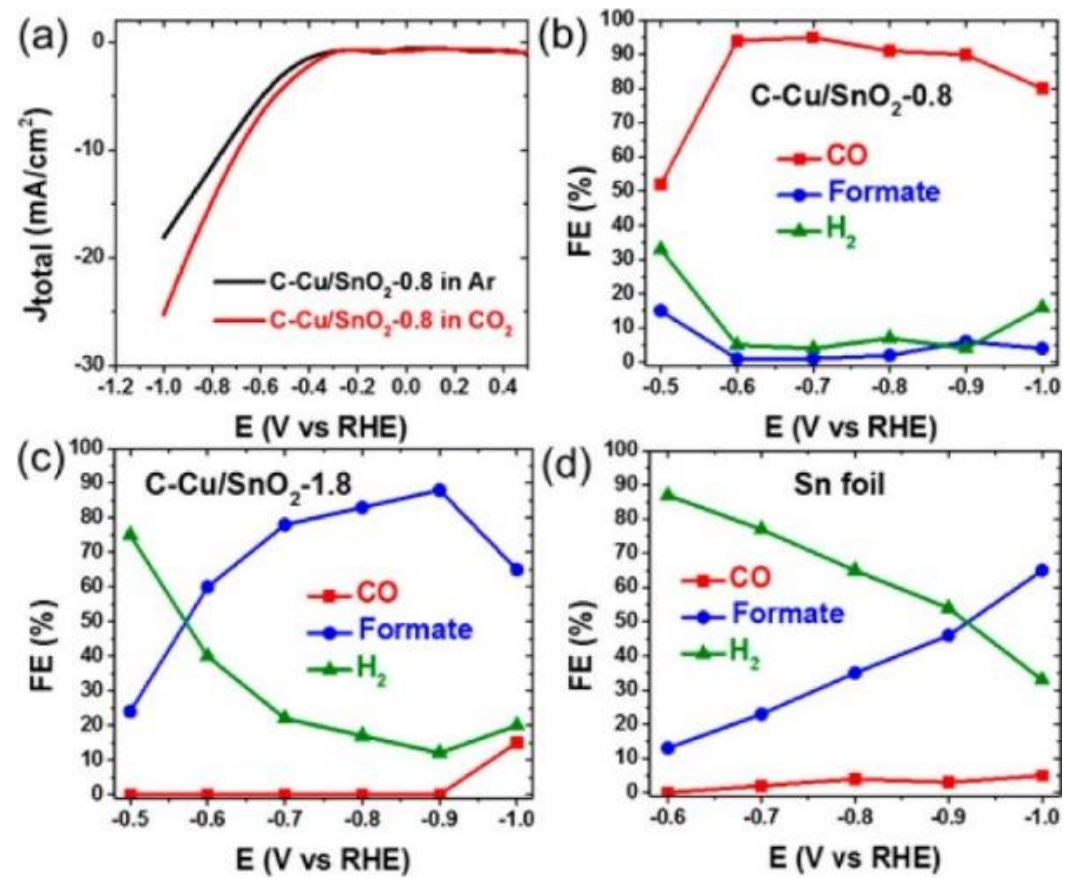

**Figure 4**. (a) (Ar- and $CO_2$-saturated LSV scans with C-Cu/$SnO_2$ ($SnO_2$ thickness 0.8 nm) catalyst in 0.5 M KHCO3 aqueous solution, Reduction potential dependent FE's for $CO_2$RR measured on (b) C-Cu/$SnO_2$-0.8 catalyst, (c) C-Cu/$SnO_2$-1.8 catalyst and (d) acid-treated Sn foil electrode. Reprinted with permissions from ref. 55. Copyright (2017) American Chemical Society.

Theoretical calculations have shed light on the concurrent alloying of $SnO_2$ with Cu, caused by the compression of the oxide shell and ultimately modifying the electrocatalyst selectivity. The same concept has been employed for driving the selective HCOOH formation (FE 80%) by Ag-Sn bimetallic core covered by a shell of $SnO_x$, the former acting as a high electron conductor, the latter as the catalytic phase. The catalyst was prepared by galvanic displacement which could permit to optimize the $SnO_x$ thickness (1.7 nm) thus tuning the $CO_2$RR performance.[54]

The core@shell structure could guarantee higher electroactive surface area as compared to 2D layered catalysts, resulting in high current densities, while the presence of the metallic core mitigates the insulating nature of the metal oxide phase for efficient charge transport. Seed-mediated approaches have proved to be a versatile tool for controlling the oxide thickness, as proved in the assembly of Cu@$In_2O_3$ NPs, where the tunable structure afforded production of syngas in various $H_2$/CO ratios.[56] Computational studies on metal/metal oxide interfaces have shown that cooperative interfacial interaction could suppress the HER process while stabilizing the key intermediates for $CO_2$RR product. Indeed, for a core-shell morphology, such cooperation is geometrically maximized. For example, DFT analysis confirmed the electron transfer from Ag to $SnO_x$ in Ag@$SnO_x$ core-shell nanoparticles, together with stabilization of key intermediates for both the $CO_2 \rightarrow CO$ and $CO_2 \rightarrow HCOOH$ thanks to dual-site cooperative binding. As a result, formation of CO readily proceeds on the Ag surface while HCOOH is formed on $SnO_x$, while the kinetic barrier to $H_2$ evolution is considerably increased (Figure 5).[57]

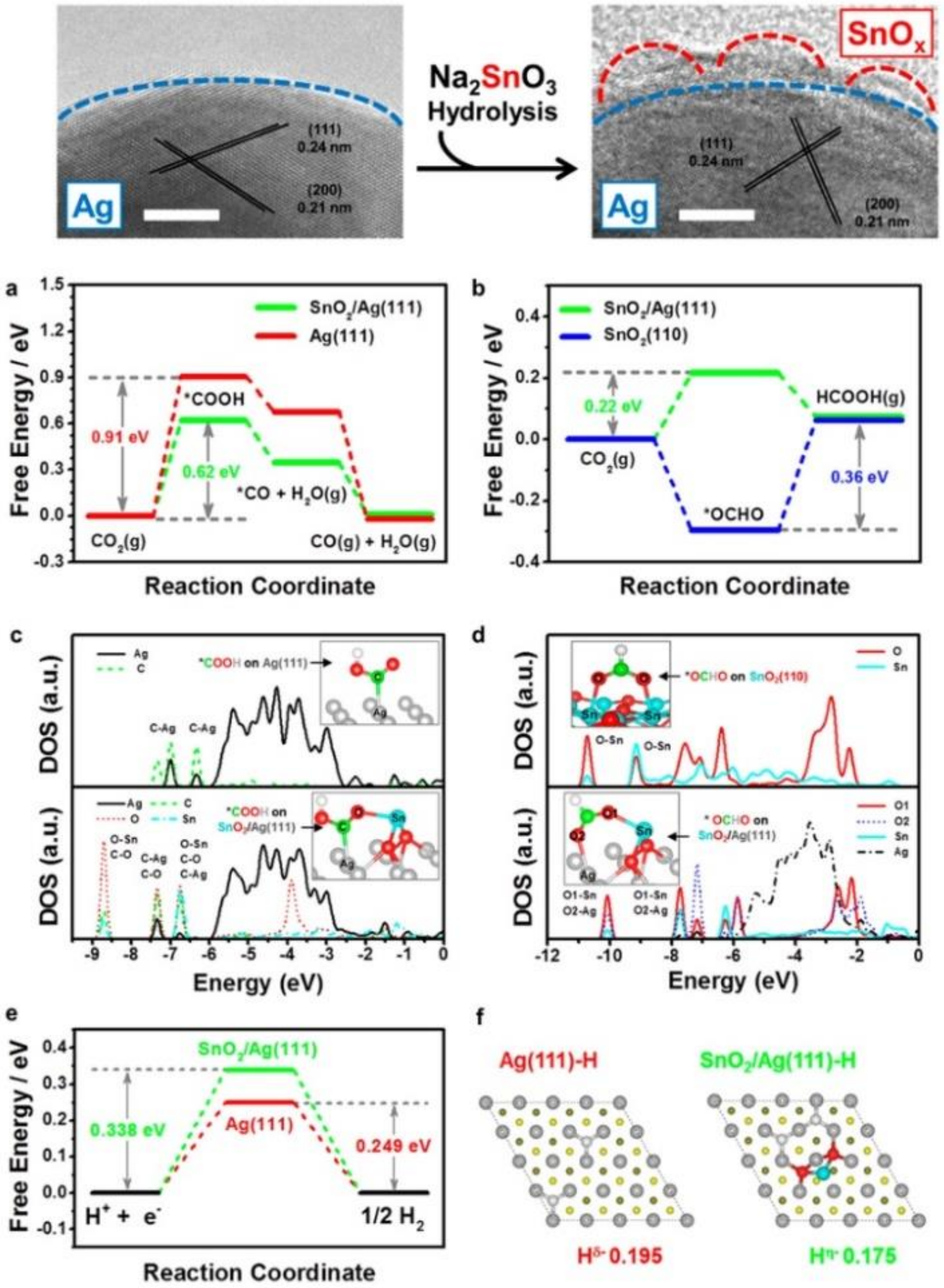

**Figure 5.** Top: TEM micrograph showing bare Ag NPs, and the Ag/$SnO_x$ nanohybrid following synthetic procedure. Bottom: Computational analysis, displaying free energy diagrams of the $CO_2$ to CO on Ag and $SnO_2$/Ag (a) and $CO_2$ to HCOOH on $SnO_2$ and $SnO_2$/Ag (b), density of states (DOS) analysis of (c) *COOH on Ag and $SnO_2$/Ag and (d) *OCHO on $SnO_2$ and $SnO_2$/Ag, with the insets showing the optimized binding geometries, (e) calculated $\Delta G_{H*}$ on Ag and $SnO_2$/Ag (e), and top views of the Ag and $SnO_2$/Ag surfaces with *H adsorbed (f). Adapted with permission from ref. 57. Copyright (2018) American Chemical Society.

A powerful strategy to improve the electrocatalyst conductivity and surface area is to combine the metal oxides with carbon supports. As a direct consequence, the electrochemical interface consists of a triple phase contact, including the metal-oxide surface, the carbon support, and the liquid electrolyte. Moreover, the introduction of the organic hetero-junction opens new opportunities vis-à-vis the control of the $CO_2$RR selectivity.

**The role of carbon nanostructures**

A current frontier in the design of a multi-phase hybrid catalysts is represented by the incorporation of carbon nanostructures (CNS) .[58] As compared to conventional carbon supports (amorphous carbons such as carbon black), CNS offer specific advantages due to their distinct textural, mechanical, electronic properties as well as tunable topography. In electrocatalysis, the fine-tuning of the interfacial CNS/metal domains is required to take full advantage of such properties. However, the understanding of the electronic properties at CNS/metal or metal-oxide interface is severely hampered by the CNS heterogeneity, which in addition to shape and size variance, also bear a large distribution of defects and surface groups, all playing a possible role in $CO_2$RR.[59] Notable attempts in correlating $CO_2$RR activity and selectivity with the carbon/inorganic interfacial characteristics rely on the combination of advanced characterization techniques and computational analysis. Centi et al. used electron microscopy, *operando* X-ray spectroscopy techniques and DFT simulations to unravel the origins of the high performance in C-C coupling by Fe oxy-hydroxide nanostructures supported on O- and N-doped graphitic carbon, where acetic acid evolved as a product with a FE as high as 97%.[60] In this system, the Fe redox chemistry is influenced by the carbon-based environment, depending on the nature of the heteroatom dopants and on the applied potential, as highlighted by DFT calculations. It turns out that the selective formation of $CH_3COOH$ occurs at single Fe(II) active sites present at the edge of the graphite layer.[60] This work is significant to highlight the complexity of the interface dynamics in carbon/metal hybrids, and of the carbon-phase substructure involving edges, steps, defects etc., because both aspects play a joint role to direct catalysis.

The use of 1D, 2D or 3D carbon nanostructures can template the final morphology of the hybrid nano-material,[10, 61] In this respect, 2D Graphene (G) has been widely employed for electrocatalytic applications due to the very high surface area combined with an unrivalled mobility of the charge carriers, flexibility and film robustness. Moreover, graphene-supports are known to be highly sensitive to doping and interfacial modifications, but the other side of the coin is that the resulting electrochemical response depends strongly on the graphene synthetic protocols, and therefore on the sample distribution and density of surface defects and on possible contaminants of the resulting materials.[62] It should be also considered that the expected surface area of 2D graphene supports is generally affected by self-stacking of the individual layers through extended π-π interactions. Therefore, fabrication of 3D- composites with multi-phase arrangements can offer a valuable opportunity to counteract graphene self-aggregation.

The use of graphene nanoribbons (GNR) as support for gold NPs was found to be essential for the electronic regulation of Au active sites, and one of the reasons for the enhanced performance in $CO_2$RR originated from the higher $CO_2$ uptake as compared to bare Au NPs, considering the ultramicroporosity (< 0.7 nm) and the improved electrochemical surface area (ECSA) of the GNR/NPs material.[63] A significant shift observed for the CO2RR overpotential is a direct proof of the intrinsic change of the electronic properties of the active sites as a result of the Mott-Schottky heterojunction formation.[63] Strong electronic interaction was invoked to justify the increased activity of a few-layer Sb/G nanocomposite prepared by coupled cathodic/anodic exfoliation of Sb and graphite. Such an interaction modifies the binding energies of the $CO_2RR$ intermediates, as demonstrated by the consistent decrease of the Tafel slope when passing from bulk Sb to Sb/G. This is associated to a change of the rate-limiting step with the Sb/G catalyst, which is no longer the $CO_2$ adsorption, but rather the one-electron reduction $*CO_2/*CO_2^{\cdot-}$ step.[64] An interesting opportunity for tuning $CO_2RR$ can be envisaged considering the *ad hoc* functionalization of the CNS surface with suitable organic pendants, installed with optimized synthetic protocols. This was recently demonstrated by preparing $SnO_x$ nanosheets/multi walled carbon nanotubes (MWCNT) hybrids, featuring three different types of pendant groups, namely -COOH, $-NH_2$ or -OH terminals (Figure 6).[65]

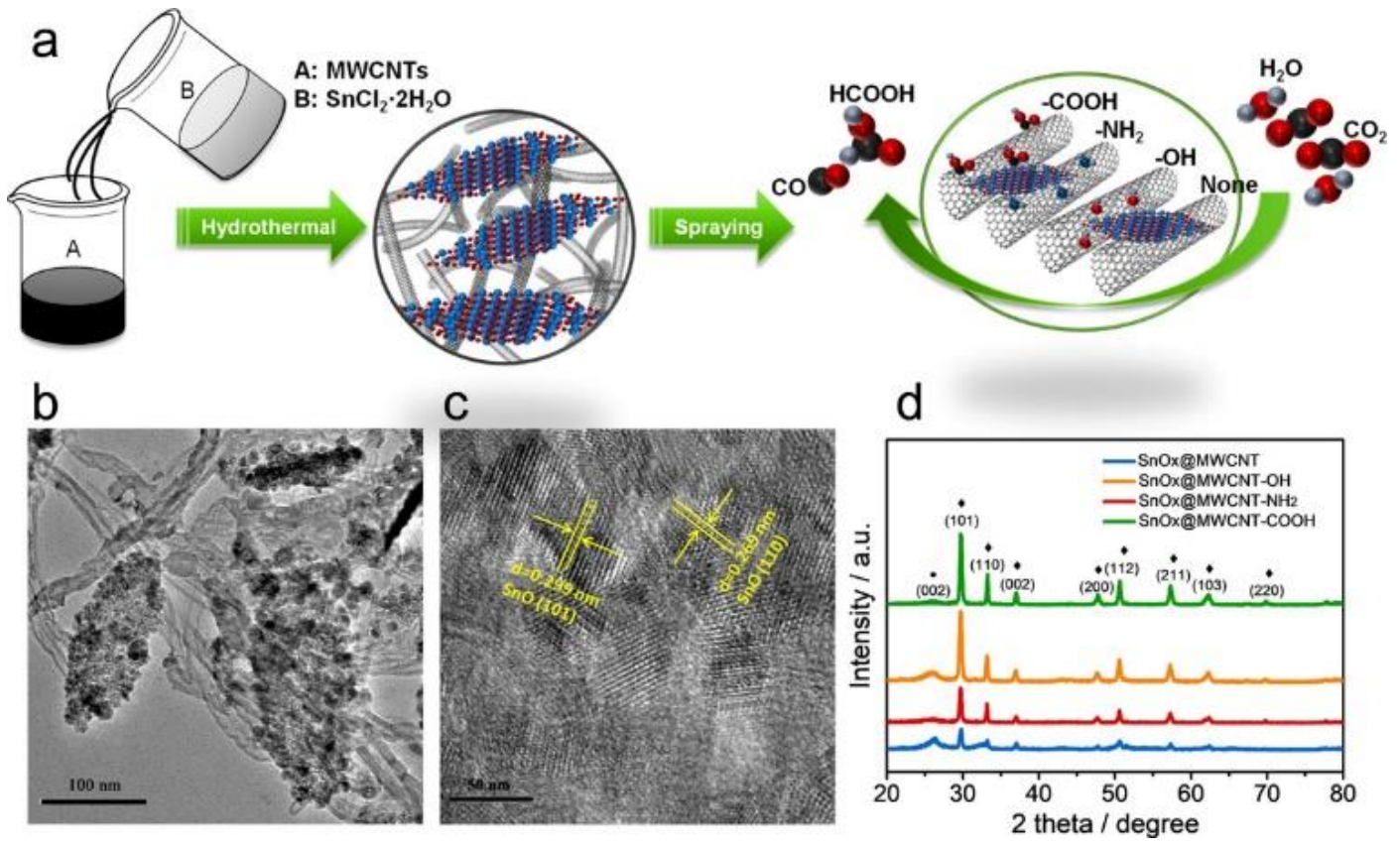

**Figure 6**. a) Sketch of the synthetic scheme for $SnO_x$/CNT catalysts, allowing uniform laying of the $SnO_x$ nanosheets, b) and c) TEM and HRTEM micrographs of the materials, d) XRD of the nanohybrids with different functional groups on the CNT. Reprinted with permission from ref. 65. Copyright Wiley and sons.

The role of the terminal group with diverse proton and electron donor properties can be traced at multiple levels: (i) determining the $SnO_x$ loading, as a function of the improved affinity of the nanocarbon surface for the metal oxide phase; (ii) increasing the available electrochemical surface area (ECSA); (iii) tuning the electronic properities of the Sn active sites by a direct inner sphere coordination effect, and by a second sphere assistance for the stabilization of reactive intermediates. Indeed, the experimental results confirmed that both the activity and the selectivity of $CO_2$RR was tuned by the MWCNT functionalization, and that $SnO_x$@MWCNT−$NH_2$ displayed an improved loading of active sites, corresponding also to an enhanced ECSA and a nearly 100 % selectivity for CO with maximal current density.[65] Further analysis on the impact of the organic domains on the $CO_2$RR selectivity needs to be addressed by drawing predictive structure-activity relationships that set the basis for a critical discussion.[66] Several examples in the literature are also highlighting the use of polymeric additives to modify the surface environment of CNS with one primary goal to boost $CO_2$RR while suppressing the competitive HER.[67, 68]

In summary the impact of CNS for $CO_2$RR elecrocatalysis can be envisaged at different levels:

(i) CNS with diverse aspect ratios and dimensionality offer a tunable platform to template the morphology of the composite electrocatalyst, tuning the surface area and porous texture

(ii) The intimate contact with the metal/metal-oxide phase provides a local modification of the active site properties including the redox state distribution, the density of defects, the hydrophobicity of the environment, electron and mass transport phenomena that can modulate the $CO_2$RR selectivity.

(iii) Alterations in crystal packing and in chemical bonding on the CNS surface can be responsible for specific activation/stabilization effects of $CO_2$RR intermediates, thus producing a unique catalytic effect.[69]

We have recently reported on a triple phase interface that is instrumental to boost electrocatalytic $CO_2$RR.[70,71] Ternary hybrids built on 3D-carbon nanohorn templates (CNH/$TiO_2$/Pd) with a hierarchical core-shell morphology, exhibited an unprecedented selectivity for formate production, at near equilibrium potential. Interestingly, Pd-assisted $CO_2$ electro-hydrogenation[27] occurred in a broad potential window, thus preventing a parallel formation of CO, which is known to poison Pd NPs, improving considerably the long term stability of the electrocatalyst.[71] Moreover, the conductive and high surface area of CNHs can facilitate electron transfer to the active sites and improve $CO_2$ mass transport versus proton diffusion, thus suppressing HER. Interestingly, $H_2$ production is associated to the reversible formate decomposition that takes place at near equilibrium potential.[71] The hierarchical design of the CNH/$TiO_2$/Pd catalyst notably allowed high activity with low loadings of the Pd precious metal, reaching a TOF of 26500 $h^{-1}$ at -0.2 V vs RHE, which sets a new benchmark in the topic.

A considerable step forward was achieved by exploiting Pd-free MWCNT/$CeO_2$ electrocatalytic interfaces for $CO_2$ reduction to formic acid. O*perando* EXAFS analysis is consistent with the involvement of transient ceria-hydride species being responsible for a direct electro-hydrogenation step. Reduction of ceria and migration of Ce(III) defects appears to be facilitated by the close contact with the conducting MWCNT surface (Figure 7).[72]

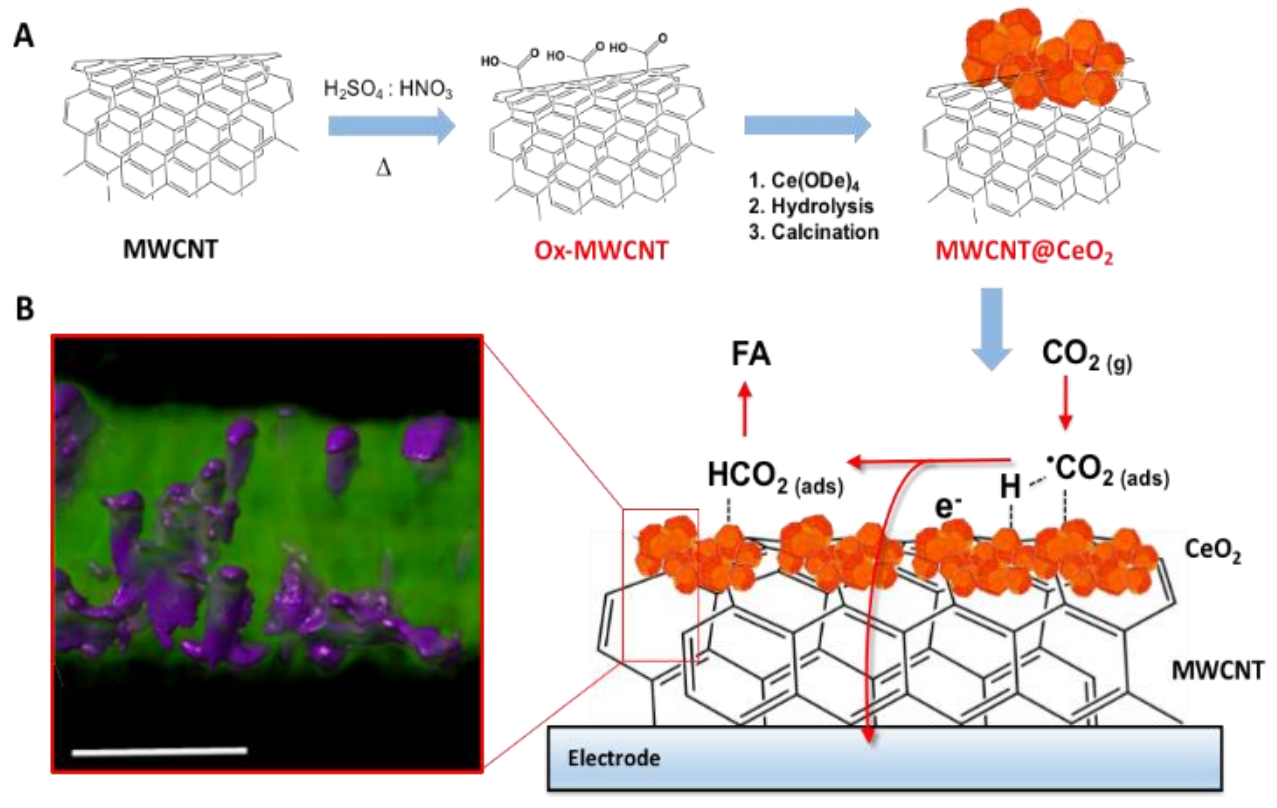

**Figure 7.** (a) schematic of MWCNT@$CeO_2$ synthesis involving a first oxidation step of the MWCNT scaffolds, followed by decoration with $CeO_2$-NPs, grown on the MWCNT surface by controlled hydrolysis of $Ce^{4+}$ tetrakis(decyloxide), $Ce(ODe)_4$, and calcination at 250 °C. (b) STEM tomographic reconstruction of **MWCNT@$CeO_2$ (**the region of high density corresponding to the $CeO_2$ is rendered with a violet mesh) and sketch of the possible mechanism of $CO_2$ hydrogenation to formic acid. Scale bar, 20 nm. Reprinted with permission from ref. 72. Copyright (2020) American Chemical Society

## Conclusions and emerging directions

The engineering of organic-inorganic hybrid interfaces can implement $CO_2$RR electrocatalysis by offering the optimal combination of efficiency, selectivity and long-term robustness. Considerable progress has been made in the field, guided by an impressive advancement of time-resolved spectroscopies and modelling studies. The fine-tuning of active site stereo-electronics by a favorable cross-talk of hybrid phase boundaries can have a formidable impact merging the gap between homogeneous and heterogeneous catalysis. Herein, we will highlight two emerging trends in anticipation of future developments:

*1) Surface engineering: controlling facets and defects at multi-phase hybrid interfaces*

When a polycrystalline material is considered, the precise identification of the structural features (step, kink, terrace, vacancy, grain boundary) that governs $CO_2$RR is perhaps impossible. Contributions from the different structural elements all sum up, and discerning priorities and synergies is a formidable challenge. One approach to address this complexity is to correlate the binding energies of $CO_2$ and of its reduction intermediates with different structural elements. Typically, single crystals are used as model catalysts to ascribe catalytically relevant structures at specific crystal facets, taking into account that crystallographic *hkl* indexing is critical for a reliable calculation of their binding mode energetics. However, because of intrinsic limitations of single crystal catalysts in terms of low current densities, the key step is to synthesize active catalysts with preferential faceting, i.e. controlling the selection of facets of metal nano-particles by suitable effectors. This research opened up a new direction for the engineering of catalytic surfaces with enhanced performance. Single crystal Cu electrodes has served as excellent examples for evaluating the importance of the crystallographic faceting for tuning $CO_2$RR selectivity, particularly in relation to C-C coupling products. Over the years, converging evidence has been collected showing that control on activity and selectivity of Cu surfaces can be attained by determining specific (*hkl*) directions for crystal growth. [73-75] General trends have been drawn for some particular facets of the Cu single crystal, in particular for the (100)-facet, which seems to favor $C_{2+}$ products, while $CH_4$ is mainly observed at the (111)-facet.[75-77] The potential-dependent selectivity of Cu(100), (111), and (751) electrocatalytic thin films prepared by physical vapor deposition (PVD) was investigated by *in situ* electrochemical scanning tunnelling microscopy, and revealed that under-coordinated active sites lead to higher selectivity towards C–C coupling products, while with Cu(751) the oxygenate/hydrocarbon product ratio was the highest.[78] These results highlight the importance of the crystal growth orientation on suitable extended interfaces, thus paralleling the results obtained with small-format single crystals**.**[78] Very recently, Cu single crystals with various morphologies and faceting have been used for the fabrication of Gas-Diffusion-Electrodes (GDE), in electrochemical cells where the observed current densities are significantly higher.[79] However a game-changer approach would be to direct the growth of active facets by a suitable choice of the hybrid interface contact as demonstrated in the case of graphene-based nanomaterials.[80]

In the realm of structural-activity relationships, one key aspect is the introduction of defects with “ad-hoc” distribution and morphology. For $CO_2$RR, the occurrence of surface defects, i.e. atom vacancies, can modulate the $CO_2$ adsorption and the binding energies of emerging intermediates, which results in a change of activity and selectivity.[81] It was for example demonstrated that a defect-rich $Bi_2S_3$-$Bi_2O_3$@rGO nanohybrid interface is determinant for the $CO_2$ adsorption, but requires to be appropriately tuned as a too high co-localization of vacancies may result in fragility, deterioration and conductivity loss.[82] In particular, the ability of oxygen vacancies to lower the activation energy barrier for stabilization of $HCOO^{-}$* intermediate was recently demonstrated for $Co_3O_4$ layered catalysts.[52] ZnO nanosheets displayed a $CO_2$ to CO activity that proportionally increased with the content of O vacancies,[51] while oxygen vacancies in Cu/$CeO_2$ was instrumental for accessing catalytic sites for selective reduction to methane.[83]

The formation of O vacancies in metal oxides can be generated through several pathways such as reduction with $H_2$ or with other chemical reductants (i.e. $NaBH_4$), thermal treatments, plasma-assisted methods, ultrasonication and others, and some of these methods can be also extended to other non-oxide materials such as dichalcogenides or nitrides, where S, Se or N vacancies have been generated.[84-86] With this aim, hybrid nanocomposites integrating carbon nanostructures can leverage a highly efficient interfacial charge transfer under electrocatalytic regime, thus providing a favorable shaping of the metal-oxide phase defects thus boosting $CO_2$RR (Figure 8).[72]

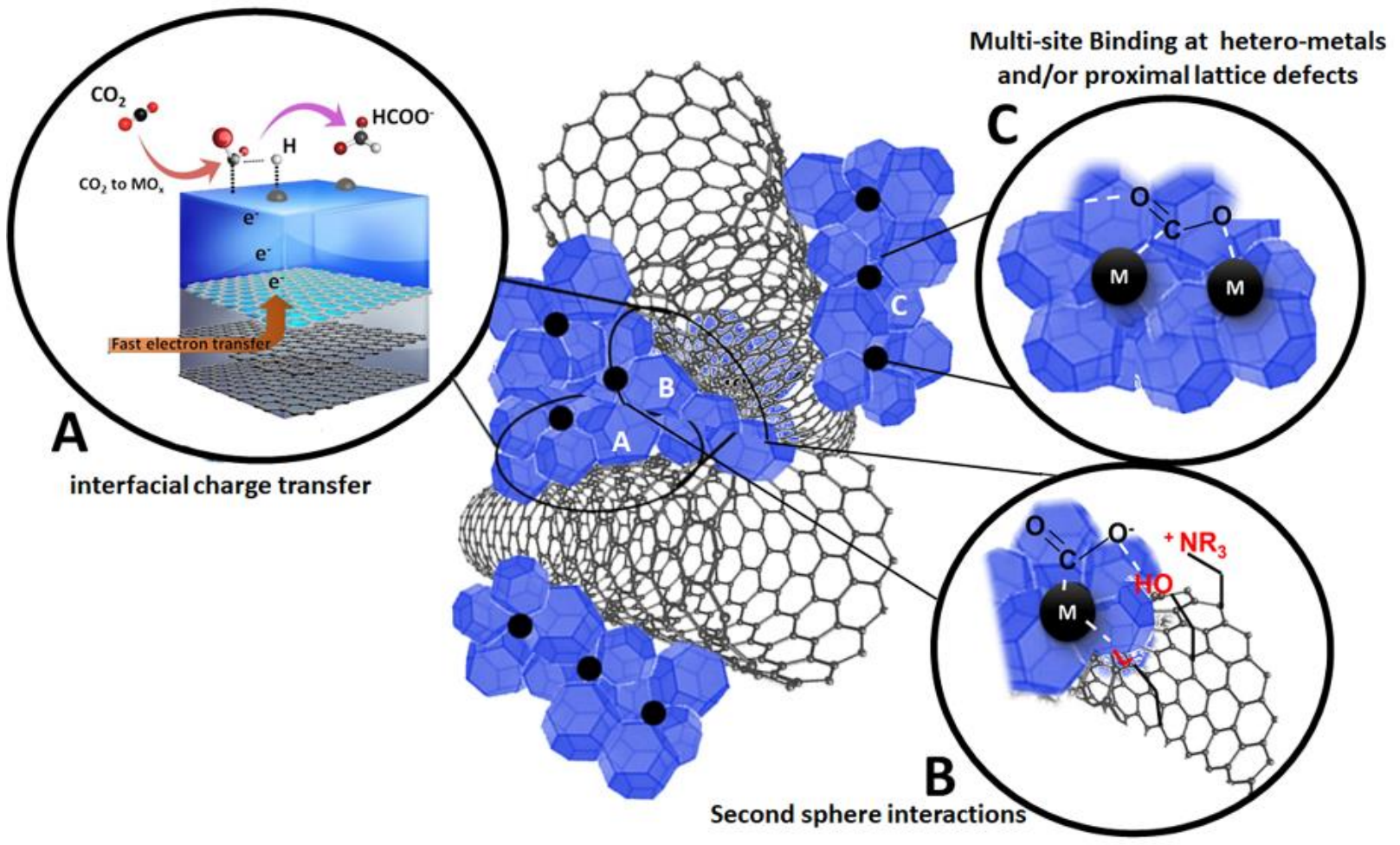

**Figure 8.** Interfacial effects occurring at multi-phase hybrid electrocatalysts. A: improved electron transfer at conductive $MO_x$/CNS interface; B: second sphere interactions promoted by terminal groups on functionalized CNS; C: multi-site binding at M/$MO_x$ interfaces.

2) *Bio-inspired catalyst design: shaping cooperative and cascade mechanisms at multi-phase hybrid interfaces*

Inspiration from natural born catalysts, i.e. enzymes, has been one priority mission of biomimetic inorganic chemistry, with the twofold aim of (i) providing a better understanding of biosynthetic pathways and (ii) discovering new catalytic manifolds with exceptional selectivity and efficiency rivalling the biological systems within artificial environments. [87] Major breakthroughs have been reported in the field of homogeneous catalysis and functional molecular systems that can be designed to replicate bio-inspired mechanistic features. The same vision translated into heterogeneous surfaces, bulk materials and hybrid nano-composites is now considered one emerging research direction with great appeal for electrocatalytic large-scale production and device exploitation.[88] Indeed, Nature has adopted a most effective task-separation, modular approach to orchestrate multiple-functions by making extensive use of interfaces and hybrid organic-inorganic domains for biological $CO_2$ processing. Natural enzymes such as carbon monoxide dehydrogenase (CODH) and formate dehydrogenase (FDH) can interconvert $CO_2$, CO, and formate under mild conditions at equilibrium potential. These enzymes exploit a synergy of effects resulting from tailored hydrophobic/hydrophilic protein domains, a "hard-soft" metal coordination environment and multi-site electron and proton transfer pathways, which are tuned by specific second-sphere and long-range stereo-electronic effectors. This strategy can be ideally transferred to the fabrication of organic-inorganic, multi-phase electrocatalytic platforms, shaped along bio-inspired guidelines but using totally synthetic building blocks.[71] In particular the combined use of metal/metal oxide domains and carbon nanostructures offers a wide space to explore the impact of the first and second sphere effects on the electrocatalytic active sites. This implies a tailored engineering of the interfacial chemistry at the molecular scale including the positioning of**:** (i) hetero-metals and/or proximal lattice defects; (ii) localized charges; (iii) proton donor/acceptor groups; (iv) spacers and/or sterically orienting groups (Figure 8). Moreover, organic additives or surface coatings have been found to enhance the electrocatalyst performance, selectivity and long-term stability by virtue of modifications of surface sites and their binding properties of specific intermediates.[89]

Two main aspects will be instrumental to leverage bio-inspired $CO_2$RR at nanohybrid platforms: (i) low-energy proton-coupled electron transfer (PCET) mechanisms and (ii) sequential catalytic steps that maximize product selectivity (tandem catalysis). In both cases, a favorable interplay of organic and inorganic interfaces can be expected. In PCET, concerted electron and proton transport can be envisaged at metal/metal oxide contacts with conductive carbon nanostructures by installation of proton acceptors/donors with tailored thermodynamic strength (pKa) so to enable multi-site electron and proton transfer events and facilitate $CO_2$RR at low overpotential. This will enhance current efficiency, while broadening the $CO_2$RR selectivity window.[90]

Tandem catalysis by multiple enzymes, that proceeds in sequential metabolic steps, is the biological way for the continuous fixation of $CO_2$ and its conversion into high-value multicarbon products.[91] Along the same concept, a cascade of electrocatalytic steps can be programmed at distal sites on the hybrid multi-phase platform, where $CO_2$RR intermediates are sequentially converted or coupled to increase the complexity of the carbon-based products. The multi-phase composite can differentiate the reactive steps by a stringent confinement of the active sites in the diverse nano-domains, while favoring the interfacial transport of reagents and the release of reactive intermediates.[92]

This synthetic scheme will require a proper choice of the different catalytic sub-units, their distribution, coverage density and interfacial connection mediated by the nanocarbon scaffolds. Electrocatalytic analogs of enzymatic cascades pose some formidable challenges with respect to the orchestration of rates and reagent/product diffusion, although retaining great potential for synthetic applications.[93]

**Declaration of Competing Interest**

The authors declare that they have no known competing financial interests or personal relationships that could have appeared to influence the work reported in this paper.

**Acknowledgements**

M. P. is the AXA Chair for Bionanotechnology (2016–2023). This work was supported by the University of Trieste, INSTM, the European Commission (H2020 – RIA-CE-NMBP-25 Program, Grant No. 862030), the European Commission (H2020 – LC-SC3-2019-NZE-RES-CC – Grant no. 884444) and the Italian Ministry of Education MIUR (cofin Prot. 2017PBXPN4).

Part of this work was performed under the Maria de Maeztu Units of Excellence Program from the Spanish State Research Agency Grant No. MDM-2017-0720.